# Fraton Theory and Modelling of Self-Assembling of Complex Structures

*M. Lavrskyi and H. Zapolsky*

*GPM, UMR 6634, University of Rouen, Saint-Etienne du Rouvray 76801, France*

*A.G. Khachaturyan\**

*Dept. of Materials Science & Eng., Rutgers University,607 Taylor Road, Piscataway, NJ 08854, USA and Dept. of Materials Science & Eng., University of California, Berkeley, Berkeley, CA 94720*

**e-mail :** khach@jove.rutgers.edu

A self-organization is an universal phenomenon in nature and, in particular, is highly important in materials systems and biology. We proposed a new theory that allowed us to model the most challenging cases of atomic self-assembling whose complexity prevented their modeling before. For example, the most challenging and biologically relevant case of formation of double-stranded helix polymers from a solution of monomers is successfully simulated. The self-organization is in the atomic scale resolution while a time resolution is commensurate with the typical diffusion time. These advancements are achieved due to introduction of two novel concepts, atomic fragments (fraton) regarded as interacting pseudo-particles and structural clusters that are central for the proposed construction of the model Hamiltonian as a bilinear expansion in structural clusters. Both novelties provide a self-organization of even disordered atomic distribution to a desired atomic structure of practically any complexity. Several other examples including a crystallization of the diamond and zinc-blende structures are presented.

**Introduction**

The self-organization in a system consisting of interacting elements is among the most fascinating phenomena. Examples can be found in almost every field of current scientific interest, ranging from coherent pattern formation in physical and chemical systems[1–8] to the various morphogenetic problems in biology[9–11] and many examples of self-organization in the Nature (solar granulation, sand dunes on Mars, vortex formations such as the Large Red Spot on Jupiter and the structure of the planetary rings, …).[12,13] In the last decades, computational modeling has provided an important tool for understanding pattern formation dynamics and self-organization in many systems.[14–16] Formation of highly complex atomic structures with special physical properties is an extremely important particular case of this problem. In the last 25 years, significant progress has been made by using molecular dynamic (MD) simulations of biological macromolecules[17]. However, at present, there are still significant difficulties to prototype the atomic self-organization of complex structures for comparatively large number of atoms if it starts from initially disordered distribution and develops during the time commensurate with the typical time of diffusion. This time scale may potentially be within a range between a fraction of seconds and years.

To address this problem, we introduced two novel concepts. They are atomic pseudo-particles, which are atomic fragments that for brevity we call fratons, and a new form of the model Hamiltonian. The proposed statistical mechanical theory of fratons allowed us to model the spontaneous self-assembly of a disordered atomic distribution into chosen in advance 3D atomic structures of seemingly arbitrary geometric complexity.

To illustrate a potentiality of this approach, we successfully modeled the self-assembly of a random system into complex atomic configurations ranging from single-component and multi-component crystals to double-helix polymers mimicking DNA molecules. As far as we know, a self-assembling of disordered systems in atomic structures of such a complexity has been never modeled before.

**Model**

We assume that each atom is a sphere comprised of its finite elements, which are atomic fragments, and treat these fragments as pseudo-particles. For brevity, we call these pseudo-particles fratons. In the multi-component systems, the number of types of fratons is equal to the number of different atomic components of the system.

Unlike the conventional approach describing the configuration of a multi-atomic system by coordinates of atomic centers, the proposed theory describes atomic configurations by occupation numbers of fratons.

The fratons are considered as interacting pseudo-particles occupying sites of the computational grid playing a role of the underlying Ising lattice. In fact, this system is a non-ideal crystal lattice gas described by the Ising model. The proper choice of a model Hamiltonian describing the interaction of fratons should result in both their "condensation" into atomic spheres and the movement of the spheres into the desirable equilibrium atomic configuration driven by the spontaneous minimization of the free energy.

In the proposed fraton model, the configurational degrees of freedom are occupation numbers, $c(\mathbf{r})$, where the vector $\mathbf{r}$ describes a site of the computational grid lattice. The function $c(\mathbf{r})$ is equal to either 1 if the site at $\mathbf{r}$ is occupied by a fraton or 0 if it is vacant. In this description, the dynamical variables are occupation numbers of the fratons rather than the coordinates of centers of atomic spheres.

To avoid interpenetration of the atoms, we introduced an analogue to the Pauli exclusion principle assuming that two fratons cannot occupy the same site of the grid lattice. The assumed exclusion provides a dynamic "exchange" repulsion preventing the atomic overlap.

At finite temperature, $T$, the evolving system is described by averaging over the time-dependent ensemble. The averaging gives the occupation probability, $\rho(\boldsymbol{r},t) = <c(\boldsymbol{r},t)>_t \leq 1$, where the symbol $<...>_t$ implies averaging over the time-dependent ensemble and $t$ is time. *The occupation probability, $\rho(\boldsymbol{r},t)$, in fact, is the probability that a point $\boldsymbol{r}$ is located anywhere inside of any sphere describing an atom at the moment t.*

The temporal evolution of the density function of fratons of the multi-component system is described by the Onsager-like kinetic equation of the Atomic Density Field (ADF) theory[18]:

$$\frac{d\rho_\alpha(\mathbf{r},t)}{dt} = \sum_{\mathbf{r}'} \sum_{\beta=1}^{\beta=m} \frac{L_{\alpha\beta}(\mathbf{r}-\mathbf{r}')}{k_B T} \frac{\delta G}{\delta \rho_\beta(\mathbf{r}',t)} \qquad (1)$$

where $L(r)_{\alpha\beta}$ is the matrix of kinetic coefficients, summation is carried out over all grid sites, $\rho_\alpha(\mathbf{r},t)$ is the occupation probability of finding a fraton of the kind $\alpha$ ($\alpha$ = 1, 2, ..., m) at the site $r$, $m$ is the number of components, $k_B$ is the Boltzmann constant, $T$ is temperature, and $G$ is the non-equilibrium Gibbs free energy functional. The condition

$$\sum_{\mathbf{r}} L_{\alpha\beta}(\mathbf{r}) = 0 \qquad (2)$$

guarantees the conservation of the total number of fratons of each kind (and thus the conservation of the total volume of all corresponding atoms).

The simplest Gibbs free energy functional entering Eq. (1) is:

$$G = \frac{1}{2}\sum_{\mathbf{r},\mathbf{r'}}\sum_{\alpha=1}^{\alpha=m}\sum_{\beta=1}^{\beta=m} w_{\alpha\beta}(\mathbf{r}-\mathbf{r'})\rho_\alpha(\mathbf{r})\rho_\beta(\mathbf{r'}) + k_B T \sum_{r}\left[\sum_{\alpha=1}^{\alpha=m}\rho_\alpha(\mathbf{r})\ln\rho_\alpha(\mathbf{r}) + \left(1-\sum_{\alpha=1}^{\alpha=m}\rho_\alpha(\mathbf{r})\right)\ln\left(1-\sum_{\alpha=1}^{\alpha=m}\rho_\alpha(\mathbf{r})\right)\right] -$$
$$\sum_{\mathbf{r}}\sum_{\alpha=1}^{\alpha=m}\mu_\alpha \rho_\alpha(\mathbf{r})$$

(3)

where $w_{\alpha\beta}(\mathbf{r}-\mathbf{r'})$ is the model potential of interaction of the pair of fratons of the components $\alpha$ and $\beta$, respectively, where **r** and **r'** are coordinates of the grid sites occupied by the fratons of this pair, $\mu_\alpha$ is the chemical potential of fratons of the kind $\alpha$. Summation over *r* and *r'* in (1) and (3) is carried out over all $N_o$ sites of the computational grid lattice. The free energy (3) corresponds to the mean field approximation[19]. It is asymptotically accurate at low and high temperatures, and its accuracy asymptotically increases if the interaction radius is much greater than the distance between interacting particles. The latter condition is automatically satisfied in our case because the continuous movement of atoms can be satisfactory described only if the computational grid increment playing the role of the spacing of the Ising lattice is much smaller than the atomic radius.

Equation (3) uses the Connolly-Williams approximation[20] mapping in this case the fraton-fraton interaction into the interaction of their pairs. A chosen model potential, $w_{\alpha\beta}(\mathbf{r}-\mathbf{r'})$, describing interaction of a pair of fratons of the $\alpha$ and $\beta$ species at points, *r* and *r'*, should guarantee the self-assembly of initially randomly distributed fratons into a desirable atomic structure and properties. The Fourier transform (FT) of such a potential is:

$$\tilde{w}_{\alpha\beta}(\mathbf{k}) = \frac{1}{N_0}\sum_{\mathbf{r}} w_{\alpha\beta}(\mathbf{r}) e^{-i\mathbf{kr}} \qquad (4)$$

where the summation is carried out over all sites of the Ising lattice (grid), and the wave vector, *k*, is defined at all quasi-continuum points, *k*, of the first Brillouin zone of the computational grid, that is, at all $N_o$ the points in the k-space permitted by the periodical boundary conditions.

**Cluster Representation of Model Hamiltonian**

The most difficult part of any theory of a self-assembling is a formulation of the model potentials that would be sufficiently simple but still able to provide the formation of the desirable atomic configuration. In this paper, such model potentials are proposed. Their formulation is based on the introduced concepts of structural clusters and cluster amplitudes.
We tested their validity, in particular, for the most challenging cases wherein a complexity of the self-assembled structures has prevented their modeling by existing method. This new approach is based on the introduced concepts of structural clusters and cluster amplitudes.

A structural cluster is defined as a minimum size group of geometric points (cluster) whose size is just sufficient to fully reproduce the main topological features of the desirable final configuration of the $\alpha$-atoms. An amplitude of structural cluster of the kind $\alpha$ and is defined as:

$$\Psi_\alpha^{cltr}(\mathbf{k}) = \sum_{j_\alpha} \omega_j(\alpha,\mathbf{k}) e^{-i\mathbf{k}\mathbf{r}_{j_\alpha}} \tag{5}$$

where summation is carried out over all points of the $\alpha$-kind cluster, the index $j_a$ numbering these points, and $\omega(k)_j$ is the weight of the contribution of each point of the cluster to its amplitude. The constants $\omega(\alpha,k)_j$ is chosen to reproduce the desirable thermodynamic and mechanical properties of the simulated atomic aggregate.

With these definitions, we present the Fourier Transform of the model potential as sum of what we call the short-range and long-range interactions:

$$\tilde{w}_{\alpha\beta}(\mathbf{k}) = \lambda_1 \tilde{\theta}_\alpha(\mathbf{k})\delta_{\alpha\beta} + \lambda_2(\mathbf{k})\Psi_\alpha^{cltr}(\mathbf{k})\Psi_\beta^{cltr}(\mathbf{k})* \qquad (6)$$

The first term in Eq. (6) describes the short-range fraton-fraton pair interaction. It is responsible for the spontaneous "condensation" of fratons into atomic spheres. The function $\theta_\alpha(r)$ and $\tilde{\theta}_\alpha(k)$ in is a spherically symmetric function of $k$ schematically presented in Figs. 1a and 1b, $r_1$ being a length parameter determining the atomic radius, $\Delta r$ is a width of positive (repulsion) part of fraton-fraton interaction, $\xi$ is a height of this repulsion part. $\lambda_1$ is a constant which determines the strength of the short-range atomic interaction.

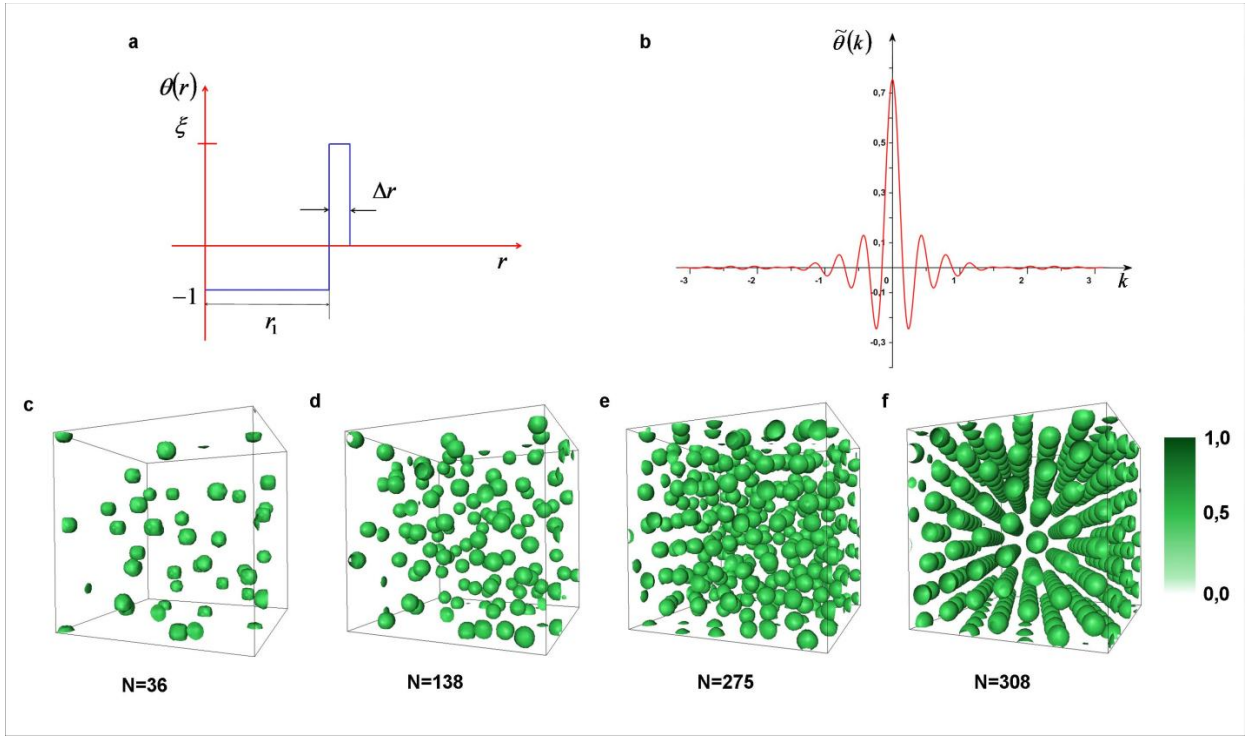

**Figure 1**. Equilibrium configurations of atomic fratons with different reduced densities self-assembled by short range interaction (first term in Eq. (7)) from disordered state. (a) Schematic representation of short-range potential $\theta(r)$. (b) Example of FT of $\theta(r)$ with the following input parameters: $\xi = 4$, $\Delta \hat{r} = 0.25$.

Final equilibrium 3D configurations of fratons for the cases with the homogeneous initial reduced with the increasing average density and input parameters: (c) $\hat{\bar{\rho}} = 0.01$ and $\hat{T} = 0.857$, (d) $\hat{\bar{\rho}} = 0.05$ and $\hat{T} = 0.913$, (e) $\hat{\bar{\rho}} = 0.1$ and $\hat{T} = 0.886$, (f) $\hat{\bar{\rho}} = 0.3$ and $\hat{T} = 0.968$. In all simulations, $\Delta \hat{r} = 0.25$; $\hat{D} = 1$, and $\hat{l} = 0.2$ were used. The input parameters were chosen to assure the same driving force for each fraton density. The numbers of atoms N on the images of the structures (c)-(d) are shown. The obtained structures in (c-e) describe a disordered distribution of atoms at different values of their density with a different degree of the short-range order typical for liquid; figure (d) describes the formation of the crystal at the higher atomic density.

The second term of Eq. (6) describes the long-range part of the fraton-fraton pair interaction responsible for the mutual arrangement of already formed atoms in the final pre-determined configuration. It is presented as a bilinear expansion in cluster amplitudes, $\Psi_\alpha^{clstr}(\mathbf{k})$. $\lambda_2$ in (6) is a fitting parameter determining a strength of the long-range interaction.

If a system is single component, the index, $\alpha$, can be dropped. Then Equation (6) is simplified to:

$$\widetilde{w}(\mathbf{k}) = \lambda_1 \widetilde{\theta}(\mathbf{k}) + \lambda_2(\mathbf{k}) \left| \Psi^{cltr}(\mathbf{k}) \right|^2 \tag{7}$$

Since the FT of fraton-fraton interaction is formulated for a specifically oriented cluster, this orientation automatically lifts the angular isotropy of the system.

**Simulation results**

To illustrate the versatility and efficiency of the fraton theory, we tested its application to the modelling of the self-assembly of three groups of 3D structures of increasing complexity. They are non-ideal gas/liquid, single-component crystals, two-component crystals, and a polymer with a double helix structure mimicking biological macromolecules. The modelling was carried out by numerical solution of the FT representation of the kinetic equation (1) using the semi-implicit Fourier spectral method[31].

We used the reduced parameters, and, in particular, average density, defined as $\hat{\bar{\rho}}_\alpha = \rho_\alpha^{at} \frac{4\pi R_\alpha^3}{3}$ where $\rho_\alpha^{at} = \frac{N_\alpha}{V}$ is the density of $\alpha$ atoms in the ground state, $N_\alpha$ is number of the atoms of sort $\alpha$, V is the total volume of the system, and $R_\alpha$ is the atomic radius of this atom. According to this definition, the reduced density, $\hat{\bar{\rho}}_\alpha$, is also a fraction of all computational grid sites occupied by fratons of the kind $\alpha$. The input parameter $\xi$ of the energy $\theta_\alpha(\mathbf{r})$, is, measured in units of $k_B T_o$ (see Fig. 1a), where $T_o$ is the solidification temperature. The lengths are measured in units of $r_1$, which is very close to the atomic radius; the grid lattice increment, $\hat{l}$ (the spacing of the underlying Ising lattice), is defined as a fraction of the atomic radius. The temperature $\hat{T}$ is also measured in units of $T_o$. The reduced time, $\hat{t}$, is measured in units of typical atomic migration time, $\tau_o$. The reduced kinetic coefficients, $\hat{L}(\mathbf{r})$, are measured in units of $\tau_0^{-1}$.

**Non-ideal gas**

The first example is a modeling of a condensation of a disordered distribution of fratons into the atomic spheres followed by their rearrangement during the equilibration process. In the case of a single-component system, we will omit the symbol $\alpha$. In all these cases, the initial disordered configuration was described by a sum of a homogeneous fratons density, $\hat{\bar{\rho}}$, and stochastically generated "infinitesimally small". Amplitude of this noise is $\sim \Delta\hat{\rho} = 0.001$. Figures 1c–1f show the final configurations of fratons with different reduced densities, $\hat{\bar{\rho}}$, equal to 0.01, 0.05, 0.1, and 0.3. In all these examples, the final structures obtained by self-assembling of fratons can be interpreted as a co-existence of two states, the liquid-like gas of atoms with high short-range order and the gas with no or very little correlation. This is expected state for the considered low density systems with the conserved number of fratons and, thus, conserved number of atoms. "Liquid" is visualized by atomic spheres distributed with the significantly short-range order while the gas is described by a disordered distribution of fratons that are not visible on the figure. The obtained difference in the number of obtained atoms with the increase of atomic density is a result of a changing of a ratio between equilibrium volume fractions of these two coexisting states under the atomic conservation condition.

In all these cases, we considered only the short-range interaction described by only the first term of the model potential defined by Eq.(7). The simulated diffraction patterns obtained for these structures are presented in the Fig. 2. The presence of diffuse scattering as a ring demonstrates a strong short-range order that is more typical for a liquid than a gas.

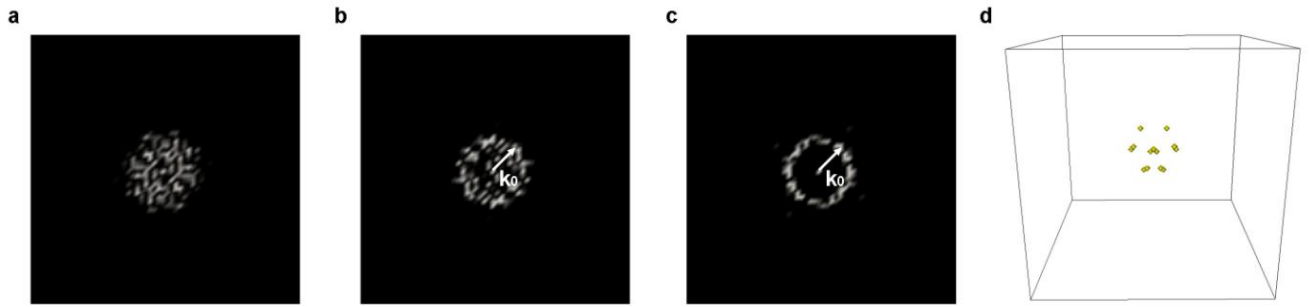

**Figure 2**. The diffraction patterns of atomic configurations formed by fratons which are shown on the Figure 1(a, b, c, d). Images (a,b,c) are (100)-plane section of the reciprocal space of their simulated diffraction patterns of the configuration on the Figure 1(a, b, c); (d) 3D image of a simulated intensity distribution in the k-space from the configuration presented on Figure 1(d).

The Fig.2 also shows that the increase of density of fratons, $\hat{\bar{\rho}}$, (and thus the increase of atomic density) results in different equilibrium structures, the structure being more ordered with the increase of the density, $\hat{\bar{\rho}}$. It is reflected in increasing of the intensity of spherical layer of intensity distribution on the diffraction patterns reflecting an increase of correlation. The latter is typical for a liquid state (Figs.1a and b). However, at $\hat{\bar{\rho}}$=0.3, the structure looks more like either imperfect crystal with the hcp or fcc close packed structure (Fig.1d) or as a two phase ( hcp+fcc) state consisting of imperfect close-packed crystalline regions and liquid with high correlation. The diffraction pattern of this configuration, shown in Fig.2d, confirms that this structure has a topological long range order.

Therefore, the chosen short-range interaction model Hamiltonian in the fraton theory, as is expected, does correctly provide a transition from the liquid/gas state to the crystalline state upon the isothermal increase of atomic density from $\hat{\bar{\rho}}$=0.01 to $\hat{\bar{\rho}}$=0.3. It also indicates that the short-range interaction (interaction radius is commensurate with the atomic diameter) reflecting the weak short-range attraction and strong repulsion of atomic cores leads to the formation of the close packed crystalline state.

Since the average atomic density and temperature are the external thermodynamic parameters of our system, the model also correctly reproduces the required change of the equilibrium state upon isothermal increase of atomic density in a single component system. This density increase results in a transition across the liquid-solid line on the $T$-$\rho$ phase diagram and produces the corresponding crystallization.

To better prototype the crystallization, we have to take into account the following. As is well known, the solidification always produces a polycrystalline state if special efforts to prevent this outcome are not made. An underlying reason for that is an isotropy of the space resulting in a degeneration of the energy of the crystalline phase with respect to any rigid-body rotation and translation. Under these circumstance, a single crystal state is achieved only if we lift this degeneration either by introducing a substrate or the external symmetry-lifting field. In our modelling, to lift this degeneration, we used the same approach to a computational prototyping of a single crystal "growth" as well: we introduced into the initial disordered distribution of fratons a layer of width $4r_1$ with the small deviations of the atomic fratons density ($\Delta\rho=0.01$) and symmetry of the fcc lattice (see Fig.3a). This layer mimics a role of a substrate.

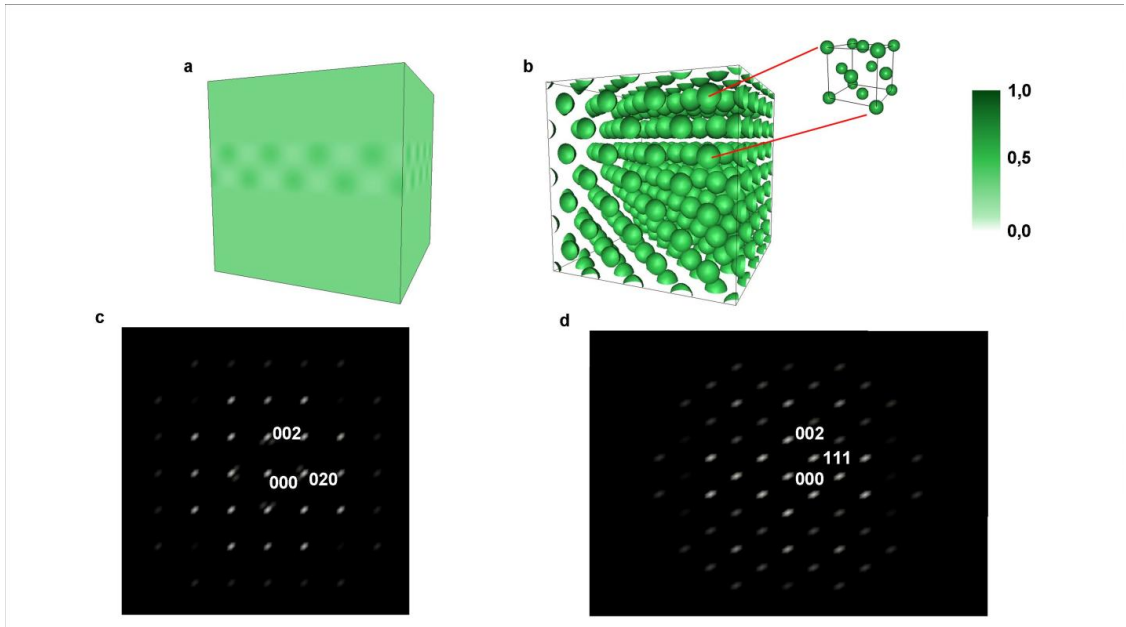

**Figure 3**. Initial (a) and final (b) configurations of fratons. The unit cell of the final structure is also shown in (b) and can be easily identified as the unit cell of the final fcc structure. The parameters in these simulations are $\hat{\lambda}_1 = $ 21.74, $\xi = 4$, $\hat{D}=1$, $\hat{\bar{\rho}}=0.25$, $\hat{l}=0.182$, $\Delta\hat{r}=0.17$, and $\hat{T}=0.924$. The size of the simulation box is 64 × 64 × 64. The initial configuration is the layer in the middle of the simulation box where small inhomogeneities ($|\Delta\rho_{max}|=1.0\cdot10^{-2}$) at the site of the fcc lattice have been introduced. In this simulation, only short-range interaction has been taken into account. The simulated diffraction pattern indicates the fcc symmetry of the atomic arrangement. The intensity distribution in the final configuration in the (100) and (111) reciprocal lattice planes are presented in (c, d), respectively.

Being "infinitesimal", these heterogeneities of the density field are too small to trigger the evolution to the "wrong" structure. However, they are sufficient to lift the translational and orientational isotropy of the continuum space, and thus to suppress the formation of a polycrystalline state. The average density in this simulation was set to be $\hat{\bar{\rho}}=0.25$.

Fig.3b shows the final equilibrium configuration of the atoms that is easily recognizes as the fcc structure. The similar conclusion can be reached analysing the diffraction patterns of this structure shown in Fig.3c and Fig.3d. The reflexions (002), (111), (220), which characterise the fcc structure, are clear seen.

**Crystallization of diamond structure**

The next level of complexity is the self-assembly of a single-component crystal with several atoms in a Bravais lattice unit cell. As an example, we have chosen a crystal with the diamond structure.

In this case, besides the short-range interaction used in the previous examples, the long-range interaction, described by the second term in Eq. (7), has been included. A chosen structural cluster in this case is a cubic unit cell of the diamond lattice: it consists of eight points in the positions of the atoms in this unit cell, as shown in Fig. 4a.

The diamond lattice has the fcc Bravais lattice with a two-atom basis, the atoms of the basis being displaced by the vector, $a(\frac{1}{4}\frac{1}{4}\frac{1}{4})$, where $a$ is the crystal lattice parameter; the coordinates are given as fractions of the crystal lattice parameter along the cube sides. In this case, besides the short-range interaction used in the previous examples, we included the long-range interaction, described by the second term in Eq. (7).

A chosen structural cluster in this case is a cubic unit cell of the diamond lattice: it consists of eight points in the positions of the atoms in this unit cell, as shown in Fig. 4a. The four points forming a fcc lattice cubic cell ( (000), $a(\frac{1}{2}\frac{1}{2}0)$, $a(0\frac{1}{2}\frac{1}{2})$, $a(\frac{1}{2}0\frac{1}{2})$ ) and additional four points obtained from them forgoing by the basis shift, $a[\frac{1}{4}\frac{1}{4}\frac{1}{4}]$.

In this case, besides the short-range interaction used in the previous examples, the long-range interaction, described by the second term in Eq. (7), has been included.

The function $\Psi^{clstr}(\mathbf{k})$ was constructed by using its definition (6) and the coordinates of the chosen structural cluster points:

$$\Psi^{clstr}(\mathbf{k}) = \left(1 + e^{-i\frac{a}{4}(k_x+k_y+k_z)}\right)\left(1 + e^{-i\frac{a}{2}(k_x+k_y)} + e^{-i\frac{a}{2}(k_y+k_z)} + e^{-i\frac{a}{2}(k_x+k_z)}\right) \tag{8}$$

where $a$ is a lattice constant of diamond structure, $k_i = \frac{2\pi m_i}{aN}$ (where $i$=x,y or z, $m_i$=1…N, N is the number of simulation grid in a given direction). With this definition, the Fourier transform of the long range interaction, $\tilde{w}_{LR}(\mathbf{k})$, can be written as:

$$\tilde{w}_{LR}(\mathbf{k}) = \lambda_2(\mathbf{k})\tilde{\Omega}_D(\mathbf{k}) \tag{9}$$

where

$$\tilde{\Omega}_D(k) = \Psi^{clstr}(\mathbf{k})\left(\Psi^{clstr}(\mathbf{k})\right)^* = \left(2 + 2\cos\left(\frac{a}{4}(k_x+k_y+k_z)\right)\right) \times$$

$$\times \left(4 + 4\left(\cos\left(\frac{k_x a}{2}\right)\cos\left(\frac{k_y a}{2}\right) + \cos\left(\frac{k_y a}{2}\right)\cos\left(\frac{k_z a}{2}\right) + \cos\left(\frac{k_x a}{2}\right)\cos\left(\frac{k_z a}{2}\right)\right)\right)$$

Then, the Fourier transform of the total fraton-fraton potentials is:

$$\tilde{w}(\mathbf{k}) = \lambda_1 \tilde{\theta}(\mathbf{k}) + \lambda_2 \tilde{\Omega}_D(\mathbf{k}) \tag{10}$$

where the first term describing the short range interaction. The functions $\tilde{\Omega}_D(\mathbf{k})$ and $\tilde{\theta}(\mathbf{k})$, were normalized by the absolute value of a difference between the maximum and minima values of these functions, respectively. The chosen parameters are $\hat{\lambda}_1 = 14.085$, $\hat{\lambda}_2 = -7.042$, $\hat{D} = 1$, $\hat{\bar{\rho}} = 0.07$, $\hat{l} = 0.286$, $\xi=2$, $\Delta\hat{r} = 0.17$, $a=4.57$, and $\hat{T} = 0.732$. The initial state is described by the randomly distributed fratons. Translational and rotational degeneracy is lifted by introducing a small static inhomogeneity, $\hat{\rho}(\mathbf{r}) = 0.03$ (it is a deviation of fraton density, $\hat{\rho}(\mathbf{r})$, from $\hat{\bar{\rho}} = 0.07$, at the points of a unit cell sites of a single unit cell of the diamond lattice), placed in the centre of the simulation box.

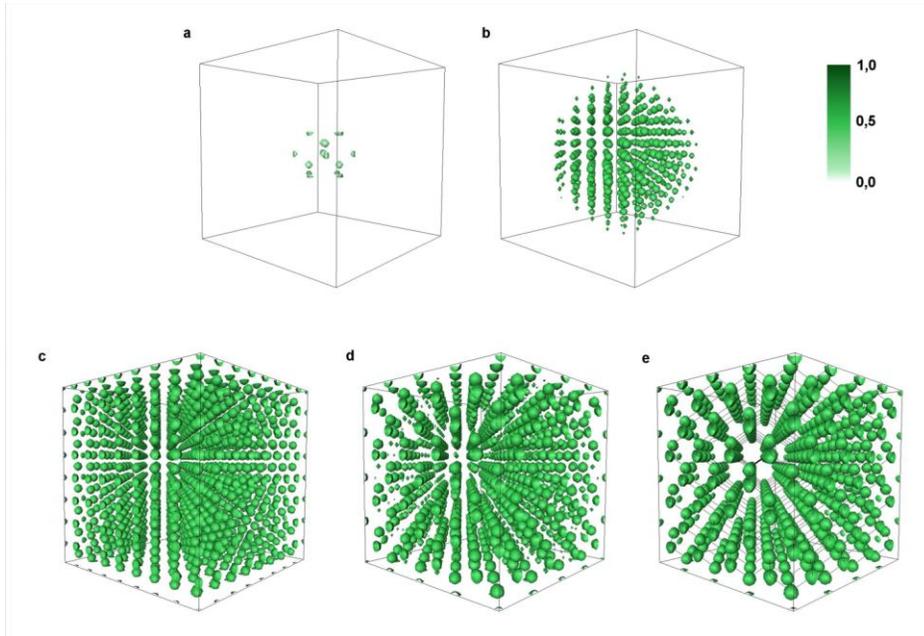

**Figure 4.** Example of a self-assembly of fratons into diamond structure at reduced times $\hat{t}$ of (a) $\hat{t} = 0$, (b) $\hat{t} = 60000$, (c) $\hat{t} = 100000$, (d) $\hat{t} = 280000$, and (e) $\hat{t} = 300000$. The parameters in this simulations are $\hat{\lambda}_1 = 14.085$, $\hat{\lambda}_2 = -7.042$, $\hat{a} = 4.57$, $\xi=2$, $\hat{D} = 1$, $\hat{\bar{\rho}} = 0.07$, $\hat{l} = 0.286$, $\Delta\hat{r} = 0.17$, and $\hat{T} = 0.732$. The initial configuration was an embryo consisting of the small variation of the fratons' density at the sites of the structural cluster of diamond structure embedded in the gas of disordered fratons. This initial configuration is the atomic cluster of the diamond structure placed in the centre of the simulation box. The size of the simulation box is 64 × 64 × 64.

The spontaneous self-organization of fratons into the diamond structure is shown in Fig. 4. The intermediate structure in the pattern formation dynamics at the reduced time t = 60000 is shown in Fig. 4b. A very interesting aspect of this self-assembling of the diamond crystal passes through is the development of the transient cubic body centered (bcc) structure (Fig. 4c) at the early stages of evolution. The lattice parameter of this bcc structure is half that of the diamond structure. This transient state gradually transformed to the diamond structure by a is followed by the gradual disappearance of some atoms in the bcc structure and the formation of the diamond structure. To better visualize the final structure, presented in Fig. 4e, the unit cell as well as the links between the first neighbours are shown.

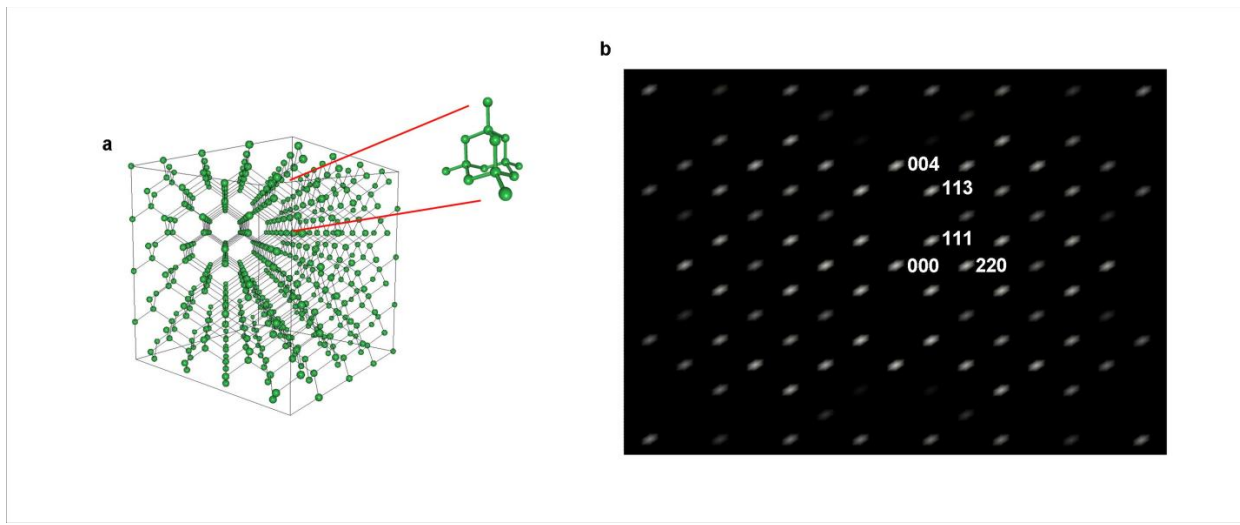

**Figure 5.** Simulated diamond structure: (a) Distribution of fratons at $\hat{t}$ = 3000 000. To clarify the structure, the links between first neighbours are drawn. (b) The diffraction pattern in the (110)-plane of the reciprocal k-space). The main diffraction peaks of the diamond structure are indicated: (111), (220), (113), and (400).

The final diamond structure and their diffraction pattern are shown in Fig.5. Fig.5 (b) shows the simulated intensity distribution in the (110) section of the reciprocal space that generated by the structure shown in Fig. 5a. This diffraction pattern is typical for the diamond structure: the most strong calculated diffraction peaks (220), (111), (311) and (400), are the strongest peaks of the diamond lattice as well. The peaks {200}, which are forbidden by the extinction rule for the diamond lattice, are also absent on the calculated diffraction pattern of the simulated structure. As is well known, the extinction rule for the diamond structure forbids reflections whose indexes are either with the indexes meeting the conditions: h, k, l are mixed odd and even; or, all even with the condition h+k+l≠4n.

**Zinc-blende structure**

Continuing our test of potentiality of the fraton theory, we gradual increased in the a complexity of the modeled structure. As example, we considered the formation of a two-component crystalline phase with the zinc-blende atomic structure.

A two-component systems atomic arrangement is formed by a "condensation" of two kinds of fratons, belonging to type A and B. This condensation should produce atoms A and B, respectively. The spontaneous arrangement caused by an equilibration of a disordered distribution of the fratons is described by the kinetic equations Eq. (1). For a two-component system, Eq. (1) in the reciprocal space is reduced to two equations:

$$\frac{\partial \tilde{\rho}_A(\mathbf{k},t)}{\partial t} = L_{AA}(\mathbf{k})\left( \tilde{w}_{AA}(\mathbf{k})\tilde{\rho}_A(\mathbf{k},t) + \tilde{w}_{AB}(\mathbf{k})\tilde{\rho}_B(\mathbf{k},t) + \left( \ln \frac{\rho_A(\mathbf{r}',t)}{1-\rho_A(\mathbf{r}',t)-\rho_B(\mathbf{r}',t)} \right)_{\mathbf{k}} \right)$$
$$+ L_{AB}(\mathbf{k})\left( \tilde{w}_{BB}(\mathbf{k})\tilde{\rho}_B(\mathbf{k},t) + \tilde{w}_{AB}(\mathbf{k})\tilde{\rho}_A(\mathbf{k},t) + \left( \ln \frac{\rho_B(\mathbf{r}',t)}{1-\rho_A(\mathbf{r}',t)-\rho_B(\mathbf{r}',t)} \right)_{\mathbf{k}} \right)$$

(11a)

$$\frac{\partial \tilde{\rho}_B(\mathbf{k},t)}{\partial t} = L_{BB}(\mathbf{k})\left(\tilde{w}_{BB}(\mathbf{k})\tilde{\rho}_B(\mathbf{k},t) + \tilde{w}_{AB}(\mathbf{k})\tilde{\rho}_A(\mathbf{k},t) + \left(\ln\frac{\rho_B(\mathbf{r}',t)}{1-\rho_A(\mathbf{r}',t)-\rho_B(\mathbf{r}',t)}\right)_\mathbf{k}\right)$$
$$+ L_{AB}(\mathbf{k})\left(\tilde{w}_{AA}(\mathbf{k})\tilde{\rho}_A(\mathbf{k},t) + \tilde{w}_{AB}(\mathbf{k})\tilde{\rho}_B(\mathbf{k},t) + \left(\ln\frac{\rho_A(\mathbf{r}',t)}{1-\rho_A(\mathbf{r}',t)-\rho_B(\mathbf{r}',t)}\right)_\mathbf{k}\right)$$
(11b)

where A and B designates two sorts of atoms and corresponding two sorts of fratons. The Fourier transforms of the interaction energies, $\tilde{w}_{\alpha\beta}(\mathbf{k})$ determined by Eq.(6), are:

$$\tilde{w}_{AA}(\mathbf{k}) = \lambda_{1A}\tilde{\theta}_A(\mathbf{k}) + \lambda_{2A}(\mathbf{k})\tilde{\Omega}_{ZB}^{AA}(\mathbf{k}) \tag{12a}$$

$$\tilde{w}_{BB}(\mathbf{k}) = \lambda_{1B}\tilde{\theta}_B(\mathbf{k}) + \lambda_{2B}(\mathbf{k})\tilde{\Omega}_{ZB}^{BB}(\mathbf{k}) \tag{12b}$$

$$\tilde{w}_{AB}(\mathbf{k}) = \lambda_{2AB}(\mathbf{k})\tilde{\Omega}_{ZB}^{AB}(\mathbf{k}) \tag{12c}$$

where $\tilde{\Omega}_{ZB}^{\alpha\beta}(\mathbf{k}) = \Psi_\alpha^{clstr}(\mathbf{k})(\Psi_\beta^{clstr}(k))^*$

The zinc-blende structure has two atoms, A and B in a primitive unit cell of the fcc Bravais lattice with positions (000) and $(\frac{1}{4}\frac{1}{4}\frac{1}{4})$, correspondingly. To describe the model Hamiltonian providing evolution of this two-component structure, we needed two structural clusters, viz., the clusters of type A and B. The cluster A consists of four points: the point (000) and the points of its nearest neighbors in the fcc lattice, $a(\frac{1}{2}\frac{1}{2}0)$, $a(0\frac{1}{2}\frac{1}{2})$, $a(\frac{1}{2}0\frac{1}{2})$. The cluster B also consists of four point. They are obtained from the four points of the cluster A by the shift, $a[\frac{1}{4}\frac{1}{4}\frac{1}{4}]$. The points of both clusters for the A and B fratons are shown in Fig.6 by green and red colors. With this definition, the cluster $\Psi$-functions for the two structural clusters are:

$$\Psi_A^{clstr}(\mathbf{k}) = \left(1 + e^{-i\frac{a}{2}(k_x+k_y)} + e^{-i\frac{a}{2}(k_y+k_z)} + e^{-i\frac{a}{2}(k_x+k_z)}\right) \tag{13a}$$

$$\Psi_B^{clstr}(\mathbf{k}) = e^{-i\frac{a}{4}(k_x+k_y+k_z)}\left(1 + e^{-i\frac{a}{2}(k_x+k_y)} + e^{-i\frac{a}{2}(k_y+k_z)} + e^{-i\frac{a}{2}(k_x+k_z)}\right) \quad (13b)$$

Using these definition the functions $\tilde{\Omega}^{\alpha\beta}{}_{ZB}(\mathbf{k})$ in the Eq.(12a-12c) can be written as:

$$\tilde{\Omega}_{ZB}^{AA}(\mathbf{k}) = \tilde{\Omega}_{ZB}^{BB}(\mathbf{k}) = 4 + 4\left(\cos\left(\frac{k_x a}{2}\right)\cos\left(\frac{k_y a}{2}\right) + \cos\left(\frac{k_y a}{2}\right)\cos\left(\frac{k_z a}{2}\right) + \cos\left(\frac{k_x a}{2}\right)\cos\left(\frac{k_z a}{2}\right)\right) \quad (14a)$$

$$\tilde{\Omega}_{ZB}^{AB}(\mathbf{k}) = 2\cos\left(\frac{a}{4}(k_x+k_y+k_z)\right) \times$$
$$\times \left(4 + 4\left(\cos\left(\frac{k_x a}{2}\right)\cos\left(\frac{k_y a}{2}\right) + \cos\left(\frac{k_y a}{2}\right)\cos\left(\frac{k_z a}{2}\right) + \cos\left(\frac{k_x a}{2}\right)\cos\left(\frac{k_z a}{2}\right)\right)\right) \quad (14b)$$

The difference in size of different species of atoms has been taken into account in the short range potential. In these simulations the ratio of two atomic radii was chosen 0.875. The size of the simulation grid, $\hat{l} = 0.25$, was measured in the unities of $r_{1A}$. Therefore, the value of $r_{1B}$ was chosen equal to $3.5\hat{l}$. For the zinc-blende structure we used the following set of input parameters: $\xi=2$, $\hat{D}=1$, $\hat{\lambda}_{1A} = 3.77$, $\hat{\lambda}_{2A} = -1.88$, $\hat{\lambda}_{1B} = 5.84$, $\hat{\lambda}_{2B} = -2.92$, $\hat{\lambda}_{2AB} = -2.26$, $\hat{l} = 0.25$, $r_{1A} = 1.143 r_{1B}$, $\Delta\hat{r} = 0.17$, $\hat{a} = 4.0$, $\hat{\bar{\rho}}_A = 0.07$, $\hat{\bar{\rho}}_B = 0.045$ and $\hat{T} = 0.235$. The initial configuration was a single unit cell embryo of the diamond embedded into the two-component gas consisting of disordered fratons of two kinds. The temporal evolution of densities of the fratons obtained by the solution of two kinetic equations (1) for two species of fratons ($\alpha$ = A, B) is shown in Fig. 6. As follows from Fig. 6e, the evolution eventually leads to the formation of the zinc-blende crystal. To better visualize the zinc-blende structure, we draw the links between the first neighbours that are also shown in Fig. 6e. In Fig. 6, green and red atoms represent the two sorts of atoms, for example Zn and S, respectively. A difference with the diamond structure is that the positions of two identical atoms in the basis of the diamond structure are occupied by different atoms in the zinc-blend structure. Each sort of atoms forms the fcc lattice with the period of the fcc Bravais lattice *a*. As is expected, these two fcc lattices are shifted with respect to each other by the distance, $a(\frac{1}{4}\frac{1}{4}\frac{1}{4})$.

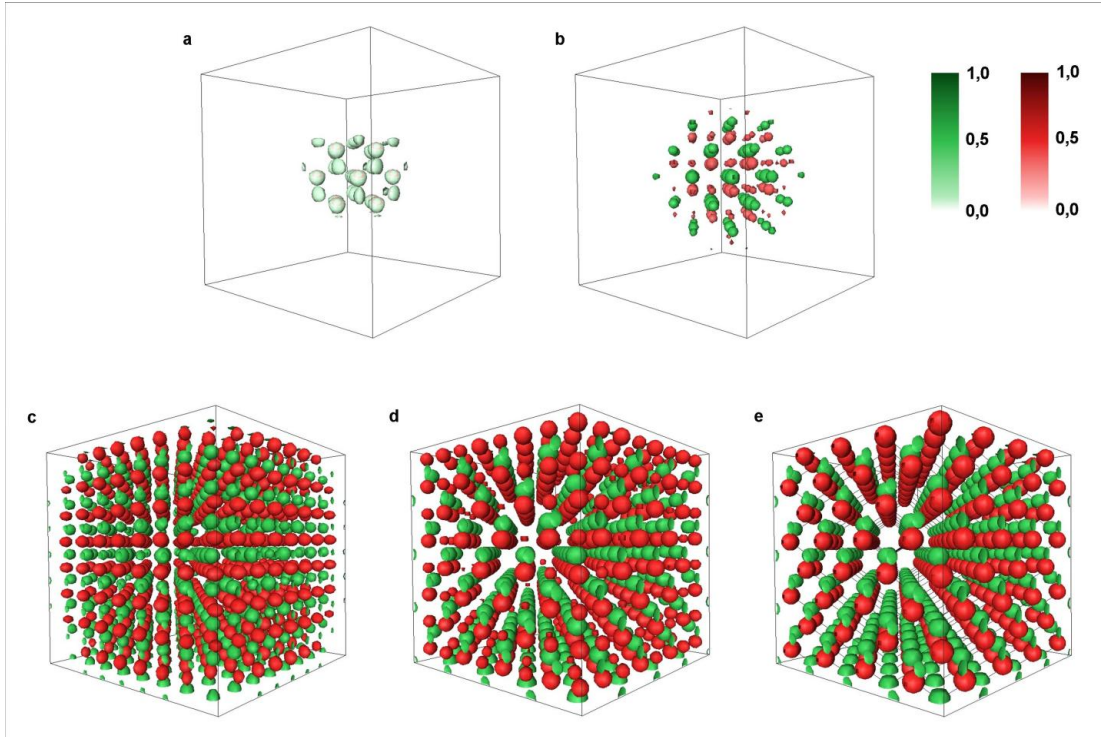

**Figure 6.** Example of a self-assembly of fratons into zinc-blende structure at reduced times $\hat{t}$ of (a) $\hat{t}=0$, (b) $\hat{t}=160000$, (c) $\hat{t}=190000$, (d) $\hat{t}=280000$, and (e) $\hat{t}=3000000$. The parameters in this simulation are $\xi=2$, $\hat{D}_{AA}=\hat{D}_{BB}=1$, $\hat{D}_{AB}=-0.5$, $\hat{\lambda}_{1A}=3.77$, $\hat{\lambda}_{2A}=-1.88$, $\hat{\lambda}_{1B}=5.84$, $\hat{\lambda}_{2B}=-2.92$, $\hat{\lambda}_{2AB}=-2.26$, $\hat{l}=0.25$, $r_{1A}=1.143 r_{1B}$, $\Delta\hat{r}=0.17$, $\hat{a}=4.0$, $\hat{\bar{\rho}}_A=0.07$, $\hat{\bar{\rho}}_B=0.045$, and $\hat{T}=0.235$. The initial configuration is the atomic cluster of a diamond structure placed in the centre of the simulation box. The size of the simulation box is 64 × 64 × 64. Two sorts of atoms with different atomic sizes are indicated in red and green.

The (110) section of the simulated diffraction pattern of the configuration shown in Fig.8a is presented in Fig.8b. The firsts strongest diffraction peaks are (200), (111),(220) and (311). The positions and intensity of these peaks are in agreement with the extinction rule: reflections from the zinc-blend structure: with mixed odd and even h,k,l indices are forbidden.

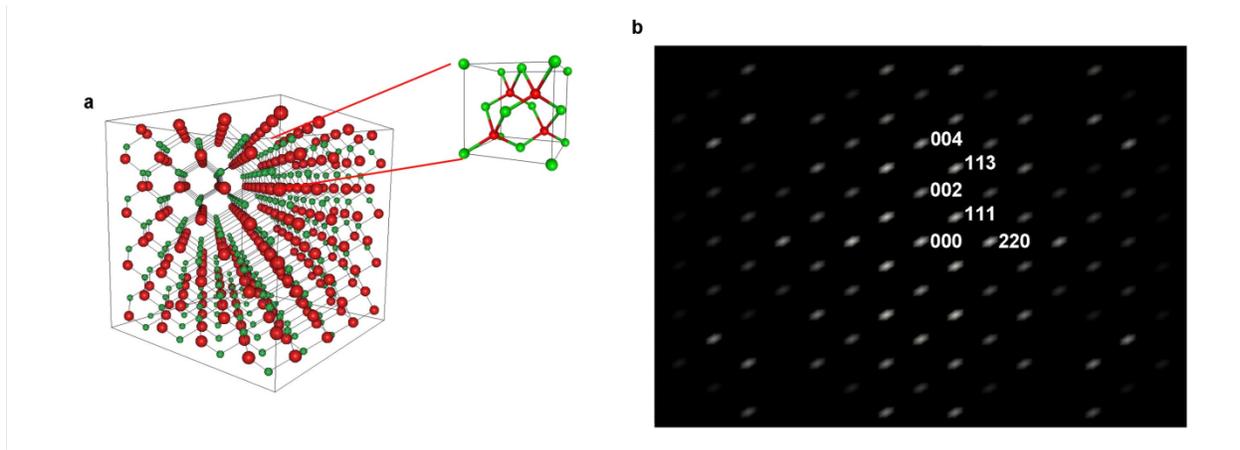

**Figure 8**. (a) Simulated configuration of fratons obtained by a spontaneous "condensation" of fratons into the atomic structure at $\hat{t}$ = 3000000. The spherical clusters of fratons of type A and B which describe atoms A and B are shown in different colors. The unit cell clarifying links between the nearest neighbours are indicated. (b) The (110) section of the reciprocal k-space. The main peaks of the zinc-blende structure are indicated: (002), (111), (311) and (400).

**Helix and double helix structures**

Molecules with a helix architecture are observed in organic materials,[21-23] helix-shaped graphite nanotubes,[24,25] liquid crystal,[26,27] proteins, and, of course, DNA and RNA polymeric molecules.[28–30] However, the most challenging test of the potency of the fraton theory would be its ability to describe a spontaneous self-assembly for the most interesting case relevant to biology, that is, the self-assembly of a double helix polymer from "soup" of randomly distributed of fratons. We chosen this system because, as far as we know, a self-assembly of randomly distributed monomers into double-tread helix polymers was a too complex phenomenon to prototype by the existing methods.

To model a spontaneous self-assembling of a helix polymer consisting of two complimentary threads, we considered two types of mutually complementary fratons whose "condensation" should produce two kinds of mutually complementary monomers.

The model fraton-fraton potential producing a helix structure should be directional and have a built-in chirality. The geometrical parameters of the configuration of the structural cluster are the pitch length, P, the number of coils per pitch, $n_0 = 6$, the distance between coils in z-direction, $h$, and the radius of the coil, $u$ (see Fig. 9).

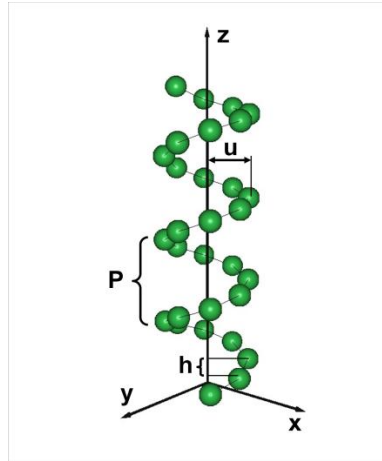

**Figure 9**. Illustration of the geometrical parameters of the helical structure: P is a pitch, $h$ is a distance between the nearest coils along the z axis and helix radius u.

Then the coordinates of points of the helix occupied by molecules are:

$$r_s = \left( u\cos\left(\frac{2\pi}{n_0}s\right), u\sin\left(\frac{2\pi}{n_0}s\right), \frac{h}{n_0}s \right) \qquad (15)$$

where s runs from 0 to $n_0$-1, $n_o$ is the number of coils in the pitch.

Our chosen structural cluster consists of two pitches. The second-pitch segment of the cluster is needed to introduce a chirality into the long-range part of the model potential. The size of the structural cluster would be drastically reduced if the chirality were built-in in the short-range part of the fraction-fraton interaction. This can be done by a straightforward modification of the short-range interaction.

Using this definition of cluster for the formulation of the function $\Psi(k)$ for the helical structure, presented in Fig.9 gives:

$$\Psi^{clstr}(\mathbf{k}) = (1 + e^{-n_0 h k_z})\left(\sum_{n=0}^{n_0-1} e^{-i(x_n k_x + y_n k_y + z_n k_z)}\right) \quad (16)$$

Then function $\tilde{\Omega}_H(\mathbf{k})$ is:

$$\tilde{\Omega}_H(\mathbf{k}) = (2 + 2\cos(h n_0 k_z))\left(6 + 2\sum_{n>m=0}^{n_0-1} \cos(\phi(n,m))\right) \quad (17a)$$

where

$$\phi(n,m) = k_x u\left(\cos\left(\frac{2\pi n}{n_0}\right) - \cos\left(\frac{2\pi m}{n_0}\right)\right) + k_y u\left(\sin\left(\frac{2\pi n}{n_0}\right) - \sin\left(\frac{2\pi m}{n_0}\right)\right) + k_z h(n-m) \quad (17b)$$

In this simulation the follow parameters were used: $\hat{D}=1$, $\hat{\lambda}_1 = 61.14$, $\hat{\lambda}_2 = -69.87$, $\hat{l} = 0.22$, $\hat{h} = \hat{u} = 1.56$, $n_0 = 6$, $\xi = 2$, $\hat{\bar{\rho}} = 0.0096$ and $\hat{T} = 0.568$. The parameters of the $\theta(k)$ function were chosen the same as for the diamond structure. The size of simulation box was 210×32×32. The initial embryo lifting the spatial and rotational energy degeneration was a one pitch inhomogeneity introduced in the centre of the simulation box. In the first step, we considered a random distribution of fratons of one kind. We also assumed that the desired helix has a single spherical monomer in each coil and approximate these monomers by a spherical shape. The latter is not a critical assumption for the theory. It is just a simplification that reduces the computational time.

The solution of Eq. (1) with the model potential given by Eqs. (17a-17b) in this case describes an evolution that eventually produces the single-stranded helix shown in Fig. 10.

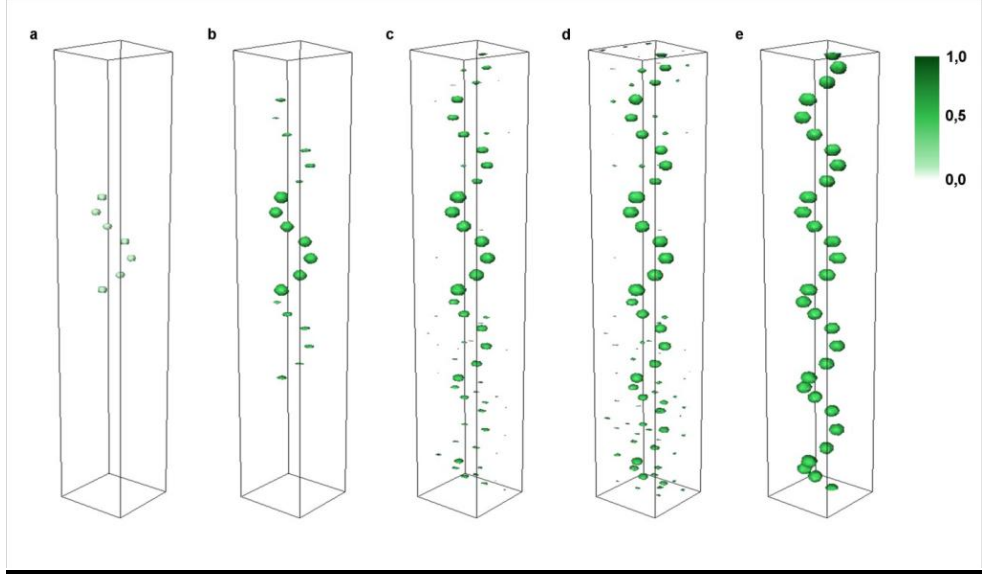

**Figure10**. Example of self-assembly of the fratons into the helix structure at reduced time of (a) $\hat{t}=0$, (b) $\hat{t}=200000$, (c) $\hat{t}=250000$, (d) $\hat{t}=300000$, and (f) $\hat{t}=700000$. The input parameters in this simulation are: $\hat{\lambda}_1 = 61.14$, $\hat{\lambda}_2 = -69.87$, $\hat{h}=\hat{u}=1.56$, $n_0 = 6$, $\xi=2$, $\hat{D}=1$, $\hat{\bar{\rho}}=0.0096$, $\hat{l}=0.22$, $\Delta\hat{r}=0.17$, and $\hat{T}=0.568$. The size of the simulation box is $32 \times 32 \times 210$. The initial configuration shown in (a) is $n_0 + 1$ coils in the helix structure.

The last example of our simulation is related to the growth of the double helix structure. This structure consists of two complimentary strands. To model a spontaneous self-assembling the double stranded helix, we had to introduced (as for the zinc-blend structure) complimentary fratons of two types, A and B, whose "condensation" should form monomers of the complementary strands. that have to form complimentary helix threads. The introduction of the second kind of fratons, which is complementary to the first kind, results in their self-assembly of a complementary strand and the formation of a double-helix molecule. The complementarity of the species is taken into consideration introducing in the model Hamiltonian a term describing the interaction of complementary fratons. We used the same type of the structural clusters for the fratons of the kind A and B. Each of them consists of two pitches of helix structure. However, the clusters are rotated with respect to each other about axis z by $\varphi=\pi$. Then, using the definition of the functions $\Psi_\alpha(\mathbf{k})$ by Eq.6 for a =A, B, gives:

$$\Psi_A^{clstr}(\mathbf{k}) = \left(1 + e^{-i\frac{n_0 h}{2}k_z}\right)\left(\sum_{n=0}^{n_0-1} e^{-i(x_n k_x + y_n k_y + z_n k_z)}\right) \quad (18a)$$

$$\Psi_B^{clstr}(\mathbf{k}) = e^{-i\frac{n_0 h}{2}k_z}\left(1 + e^{-i\frac{n_0 h}{2}k_z}\right)\left(\sum_{n=0}^{n_0-1} e^{-i(x_n k_x + y_n k_y + z_n k_z)}\right) \quad (18b)$$

We assume that $\tilde{\Omega}_{DH}^{AA}(\mathbf{k}) = \tilde{\Omega}_{DH}^{BB}(\mathbf{k}) = \tilde{\Omega}_H(\mathbf{k})$, were $\tilde{\Omega}_H(\mathbf{k})$ is defined by Eq.S14a. Then the function $\tilde{\Omega}_{DH}^{AB}(\mathbf{k})$ is:

$$\tilde{\Omega}_{DH}^{AB}(k) = 2\cos\left(h\frac{n_0}{2}k_z\right)\left(2 + 2\cos(hn_0 k_z)\right)\left(6 + 2\sum_{n>m=0}^{n_0-1}\cos(\varphi(n,m))\right) \quad (19)$$

The simulation box size and input parameters for two kinds of complimentary fratons were chosen the same as for a single-thread helix and the interaction between helix is defined by $\hat{\lambda}_{2AB} = -1.78$, $\hat{T} = 0.033$.

In spite of all these oversimplifications, this model describes some generic features relevant to the spontaneous formation of single-stranded polymeric molecule and the growth of the complementary strand of the monomers eventually producing a double-stranded helix configuration (see Fig. 11). In this case, the first single-stranded helix is a template for the aggregation on it of complementary monomers to form a double-stranded helix. For clarity, we show the clusters of fratons (monomers) of the second strand in red.

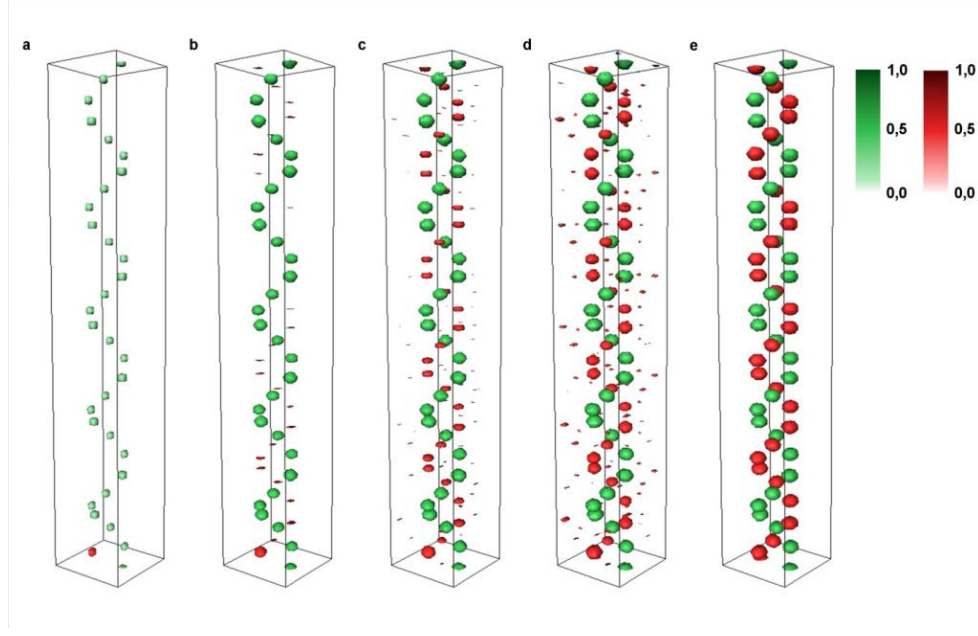

**Figures 11.** Self-assembly of the fratons into a double helix structure at reduced time of (a) $\hat{t}=0$, (b) $\hat{t}=150000$, (c) $\hat{t}=200000$, (d) $\hat{t}=300000$, and (f) $\hat{t}=1500000$. The input parameters in this simulation are $\hat{\lambda}_{1A}=\hat{\lambda}_{1B}=4.07$, $\hat{\lambda}_{2A}=\hat{\lambda}_{2B}=-4.07$, $\hat{\lambda}_{2AB}=-1.78$, $r_{1A}=r_{1B}$, $\hat{h}=\hat{u}=1.56$, $n_0=6$, $\xi=2$, $\hat{D}_{AA}=\hat{D}_{BB}=1$, $\hat{D}_{AB}=-0.5$, $\hat{\bar{\rho}}_A=\hat{\bar{\rho}}_B=0.0096$, $\hat{l}=0.22$, $\Delta\hat{r}=0.17$, and $\hat{T}=0.033$. The size of the simulation box is $32\times32\times210$. The initial configuration shown in (a) is one helix and one coil of the second helix.

## Concluding remarks

In this paper, we have chosen several examples indicating a place of the proposed theory and theory-based modeling in the existing family of available approaches to self-assembling of multiatomic systems. We selected the most difficult cases wherein the initial system is atomically disordered so that its configuration "knows" nothing about the final architecture that should spontaneously self-assembled. This self-assembling is driven only by the proposed cluster model Hamiltonian. On a top of that, we considered situations wherein the self-assembling usually takes a long time, which may range from a fraction of a second to years. A typical time of this evolution is dictated by the typical time of evolution of time-dependent ensemble rather than typical times of atomic dynamics like time of atomic vibrations. This and an absence of adequate model potentials that would describe strongly anisotropic bonding in these materials-- the problem that is solved in this paper by introducing a cluster expansion protocol for the construction of model Hamiltonian - probably, were obstacles preventing an ability of such methods as Molecular Dynamics or Monte Carlo to model a spontaneous formation of complex crystals and polymers from a liquid solution of atoms or monomers.

In fact, the developed approach opens a way to answer numerous outstanding questions concerning the atomistic mechanisms of the formation of defects (dislocations, grain boundaries, etc.), nucleation in solid–solid transformations, the formation of polymers due to aggregation of monomers in their solution, folding and crystallization of polymers, and their responses to external stimuli. This list can be significantly extended. It also allows up to study the properties of the obtained structures addressing the problem of their response to applied field.

Finally, the use of the new model Hamiltonian formulated in terms of the structural clusters and proposed fraton model provide already ready tool to address a general problem of spontaneous pattern formation by self-assembling of *any* randomly distributed building elements. These building elements can be even of the macroscopic nature.[1]

---

[1] This approach, however, can be straightforwardly extended for the prototyping of self-assembly of monomers with more complex molecular structures. In the latter case, we have to generalize the concept of fractons of atoms by introducing fracton of molecules and modify accordingly the model short-range part of the model fracton–fracton Hamiltonian. This modification should provide a "condensation" of the molecular fractons into molecules and subsequent self-assembly of these molecules.

**Acknowledgement**

We thank to R. Patte for helpful advices and discussions during code optimization. This work was supported in part by the grant from the French National Agency for the Research (ANR) project Spiderman. The simulations have been performed at the Centre de Ressources Informatiques de Haute-Normandie (CRIHAN).

Contacts information:
Pr. Armen Gourgenovich Khachaturyan
Dept. of Materials Science & Eng.,
University of California, Berkeley,
Berkeley, CA 94720
Tel.: +1 925 388 0881
E-mail address: khach@jove.rutgers.edu

Pr. Helena Zapolsky
GPM, UMR 6634, University of Rouen
Av. de l'Université, BP-12
76801 Saint-Etienne du Rouvray, France
Office : +33 232 9550 42
FAX : +33 232 9550 32
E-mail : helena.zapolsky@univ-rouen.fr

Mykola Lavrskyi
GPM, UMR 6634, University of Rouen
Av. de l'Université, BP-12
76801 Saint-Etienne du Rouvray, France
FAX : +33 232 9550 32
E-mail : mykola.lavrskyi@etu.univ-rouen.fr